\documentclass[final,5p,times,twocolumn]{elsarticle}

\usepackage{amssymb}
\usepackage{xspace}
\usepackage{txfonts}
\usepackage{graphicx}

\def \g {$\gamma$}

\journal{Astroparticle Physics}

\newcommand{\FermiLAT}{\textit{Fermi}-LAT\xspace}

\begin{document}

\begin{frontmatter}



\title{Potential of EBL and cosmology studies with the Cherenkov Telescope Array\\ 
%\StrRight{\currfilebase}{6}
}

%TODO:
%- GRB with different arrays in same plot
%- Update population study section to match survey text


\author[dm]{Daniel Mazin}
\ead{mazin@ifae.es}

\author[mr]{Martin Raue}
\ead{martin.raue@desy.de}

\author[bb]{Bagmeet Behera}

\author[si]{Susumu Inoue}

\author[yi]{Yoshiyuki Inoue}

\author[tn]{Takeshi Nakamori}

%\author[dt]{Diego F. Torres}

\author[yi]{Tomonori Totani}

\author{for the CTA Collaboration}

\address[dm]{Institut de Fisica d'Altes Energies (IFAE), Edifici Cn. Universitat Autonoma de Barcelona, E-08193 Bellaterra (Barcelona), Spain}

\address[mr]{Institut f\"ur Experimentalphysik, Universit\"at Hamburg, D-22761 Hamburg, Germany}

\address[bb]{DESY, Platanenallee 6, 15738 Zeuthen, Germany}

\address[si]{Institute for Cosmic Ray Research, The University of Tokyo, Kashiwa, Chiba 277-8582, Japan}

\address[yi]{Department of Astronomy, Kyoto University, Sakyo-ku, Kyoto 606-8502, Japan}

\address[tn]{Research Institute for Science and Engineering, Waseda University, Shinjuku, Tokyo 169-8555, Japan}

%\address[dt]{??}

%\address[tt]{Department of Astronomy, Kyoto University, Sakyo-ku, Kyoto 606-8502, Japan}

\begin{abstract}

Very high energy (VHE, E $>$100\,GeV) $\gamma$-rays are absorbed via interaction with low-energy photons from the
extragalactic background light (EBL) if the involved photon energies are above the threshold for electron-positron pair creation. 
The VHE gamma-ray absorption, which is energy dependent and increases strongly with redshift, distorts the VHE 
spectra observed from distant objects. The observed energy spectra of the AGNs carry, therefore, an imprint of the EBL. 
The detection of VHE gamma-ray spectra of distant sources (z = 0.11 - 0.54)
%by H.E.S.S. and MAGIC
by current generation Imaging Atmospheric Cherenkov Telescopes (IACTs)
enabled to set strong upper limits on the EBL density, using certain basic assumptions about blazar physics. 
In this paper it is studied how the improved sensitivity of the Cherenkov Telescope Array (CTA) and 
its enlarged energy coverage will enlarge our knowledge about the EBL and its sources.
CTA will deliver a large sample of AGN at different redshifts with detailed measured spectra. In addition, it will provide the exciting opportunity to use gamma ray bursts (GRBs) as probes for the EBL density at high redshifts.
%We point out that in the CTA era not only energy spectra of distant AGNs can be used to study EBL and cosmology, but also GRBs have a large potential for such studies.

\end{abstract}


\end{frontmatter}

% \linenumbers


%--------------------------------------------------------------------------------
% Introduction
%--------------------------------------------------------------------------------
\section{Introduction}
\label{Sec:Introduction}

\subsection{Cherenkov Telescope Array}

The proposed Cherenkov Telescope Array (CTA)\footnote{http://www.cta-observatory.org/} 
\cite{actis:2011a:cta:design,bernloehr:2012a:ctamcpaper} is a large array of
Cherenkov telescopes of different sizes for the detection of very-high energy
gamma-rays.
CTA will cover an energy range from 10s of GeV up to 100s of TeV with an
unprecedented sensitivity -- a factor of 10 or more improvement over current
generation instruments  such as H.E.S.S
\footnote{http://www.mpi-hd.mpg.de/hfm/HESS/},
MAGIC\footnote{http://wwwmagic.mpp.mpg.de/},   and
VERITAS\footnote{http://veritas.sao.arizona.edu/} --, and provide an excellent
precision in energy (order of 10--15\%), angular (down to arcmin scale) and
temporal (down to seconds) resolution \cite{actis:2011a:cta:design}.
Different layouts of CTA with different performance characteristics are
currently under investigation to balance the financial budget vs. the science return.
The different  CTA layouts and their estimated performance are described in
detail in a separate paper in this issue \cite{bernloehr:2012a:ctamcpaper}.
The present paper is focused on the projected performance of particular CTA layouts in
respect to studying the extragalactic background light and cosmology.
It is demonstrated that CTA has a large potential to provide a significant
contribution to such studies making them one of the central science drivers for
the instrument.

%--------------------------------------------------------------------------------
\subsection{The extragalactic background light}

 %------------------
% FIGURE: EBL models F1
\begin{figure*}[tbp]
\centering
\includegraphics[width=0.75\textwidth]{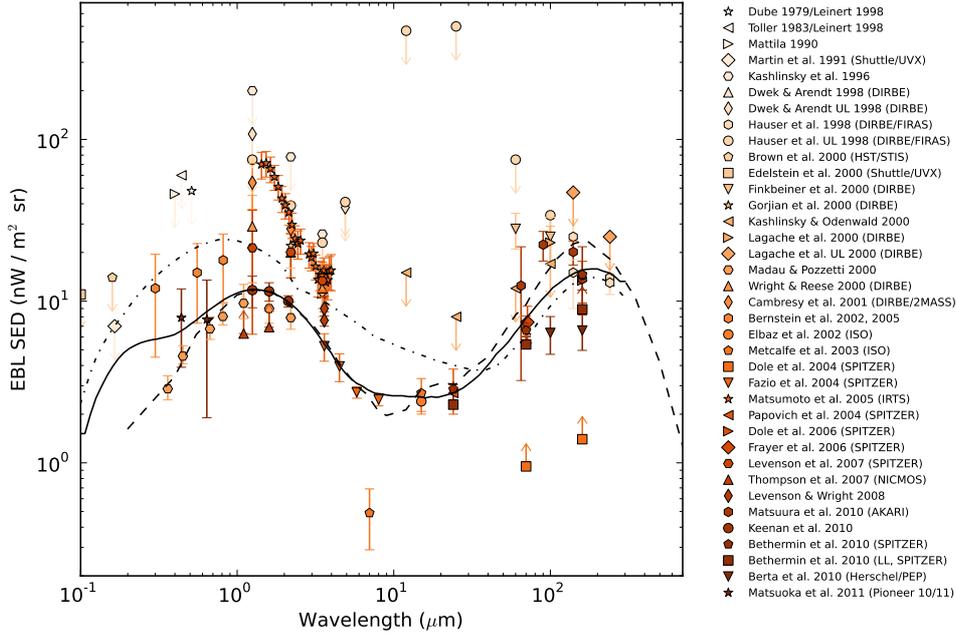}
\caption{Measurements of and limits on the EBL density in comparison to different EBL models/shapes utilized in this study at $z=0$ (dashed dotted line: fast evolution model from \cite{stecker:2006a}; solid line: \cite{raue:2008b}; dashed line: \cite{franceschini:2008a}). References for the collection of measurements can be found in \cite{mazin:2007a} and \cite{raue:2011a}.}
\label{Fig:EBLModels}
\end{figure*}
 %------------------
 
During the star and galaxy formation history diffuse extragalactic radiation in
the ultraviolet (UV)  to far-infrared wavelength (FIR) regimes was accumulated.
This radiation field, commonly referred to as the extragalactic background
light (EBL), is the second largest background after the Cosmic Microwave
Background of 2.7~K (CMB) in terms of energy contained. While the CMB conserves
the information on the structure of the universe at the moment of decoupling of matter and
radiation following the Big Bang (at redshift z $\approx$ 1000), the EBL is an
integral measure of the entire radiative energy
released dominantly by star formation that has occurred since the decoupling.
 The EBL energy
density from UV to FIR is expected to display a two peak structure: a first
peak around $\sim1\,\mu$m from integrated starlight and a second peak at $\sim
100\,\mu$m from infrared emission from dust, which reprocesses the direct
emission in the UV to optical.

% EBL measurements direct and integrated source counts.
The precise level of the EBL density is not well known (see \cite{hauser:2001a} for a review). It is difficult to measure directly due to strong foregrounds from our solar system and the Galaxy \cite{hauser:1998a}. Lower limits on the EBL density can be derived from integrated deep galaxy
counts at optical and infrared wavelengths \cite[e.g.][]{madau:2000a,fazio:2004a,frayer:2006a,dole:2006a}.
% EBL limits from VHE observations
The observation of distant sources of very high energy (VHE, typical
$E>100$\,GeV) $\gamma$-rays using Imaging Atmospheric Cherenkov Telescopes
(IACT, such as H.E.S.S., MAGIC, or VERITAS) provides a unique way to
derive indirect limits on the EBL density. This method will be discussed in
more detail in the next section. The precision of the EBL constraints set by
the IACTs improved remarkably in the last few years
(e.g. \cite{aharonian:2006:hess:ebl:nature,mazin:2007a,albert:2008:magic:3c279:science}).
Recently, data in the GeV energy range from the Fermi-LAT instrument provided new tools to derive constraints from VHE data 
(e.g. \cite{orr:2011a,meyer:2012a}).
Contemporaneously with the IACT constraints,  there has been rapid progress in
resolving a significant fraction of this background with the deep galaxy counts
at infrared wavelengths from the \textit{Spitzer} satellite (e.g.
\cite{fazio:2004a,frayer:2006a,bethermin:2010a}) as well as at sub-millimeter
wavelengths from \textit{Herschel} or the Submillimeter Common User Bolometer
Array (SCUBA) instrument (e.g. \cite{berta:2010a}). 
Furthermore, the trends of source counts at the faintest magnitudes as measured by HST and 8m-class ground-based telescopes (Keck, Subaru, VLT) indicate that the EBL contribution from standard galaxies is largely resolved  \cite{madau:2000a,totani:2001a}.
The current status of direct and indirect EBL measurements are summarized in 
an accompanying review article in this journal and can also be found in \cite{raue:2011a}.

In total, the collective limits on the EBL between the UV and FIR confirm the expected two peak structure, although the absolute level of the EBL density remains uncertain by a factor of two to ten (Fig.~\ref{Fig:EBLModels}).  In addition to this consistent picture, several direct measurements in the near IR have also been reported (e.g. \cite{matsumoto:2005a}), significantly exceeding the expectations from source counts (see \cite{hauser:2001a} and \cite{Kashlinsky2005:EBLReview} for reviews).  If this claimed excess of the EBL is real
(but see \cite{dwek:2005b, aharonian:2006:hess:ebl:nature}),
it might be attributed to emission from more exotic sources like, e.g., the first stars in the history of the universe.

% ALPS etc.
Alternative models do exist, which could, in principle, reconcile the observations with a high level of the EBL: axion like particles (ALPs), which had been introduced to address the strong-CP violation problem in particle physics, could lead to a conversion of VHE gamma-rays to ALPs. This mechanism and the related physics for CTA is discussed in a separate paper in this special issue \cite{barrio:2011a}. Recently, the secondary gamma-rays produced along the line of sight by the interactions of cosmic-ray protons with background photons have been invoked to explain the VHE spectra of distant sources \cite{essey:2011a}.

%--------------------------------------------------------------------------------
\subsection{VHE \g-rays from blazars as a probe of the EBL}
 
 %------------------
% FIGURE: EBL attenuation F2
\begin{figure}[tbp]
\centering
\includegraphics[width=0.5\textwidth]{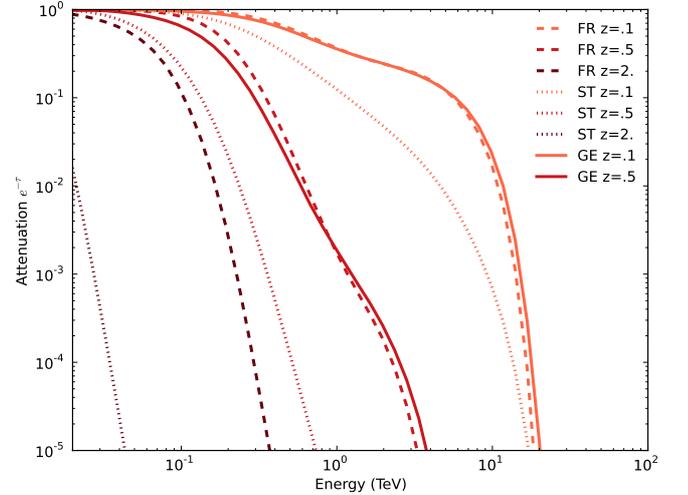}
\caption{EBL attenuation for VHE $\gamma$-rays for different EBL models and redshifts (z)  (FR: \cite{franceschini:2008a}; ST: fast evolution model from \cite{stecker:2006a}; GE: \cite{raue:2008b}). Color code of the curves is related to different redshifts (the darker the further away).}
\label{Fig:EBLAttenuation}
\end{figure}
 %------------------

On the way from the source to the observer, VHE \g-rays can suffer absorption losses by interaction with the low energy photons of the EBL via the pair-production mechanism \cite{nikishov:1962a,gould:1967a}:
\small
\setlength{\mathindent}{0pt}
\begin{equation}
  \gamma_{\mbox{\tiny{VHE}}} + \gamma_{\mbox{\tiny{EBL}}} \longrightarrow e^+ + e^- \;\;\;\; 
      \mathrm{with} \;\;E_{\gamma_{\,\mathrm{VHE}}} \cdot E_{\gamma_{\,\mathrm{EBL}}} > (m_e c^2)^2 \,.
\end{equation} 
\normalsize
\setlength{\mathindent}{2pt}

VHE \g-ray photon fluxes measured on earth $\Phi_{\rm obs}(E_{\gamma})$ are, therefore, altered from the flux emitted at the location of the source $\Phi_{\rm int}(E_{\gamma})$
\small
\setlength{\mathindent}{0pt}
\begin{equation}
\Phi_{\rm obs}(E_{\gamma}, z) =  e^{-\tau(E_{\gamma}, z)} \times \Phi_{\rm int}(E_{\gamma}) \,.
\end{equation} 
\normalsize
\setlength{\mathindent}{2pt}

The optical depth, $\tau(E_\gamma)$, encountered by VHE $\gamma$-rays of energy $E$
emitted at redshift $z$ and traveling toward the observer 
can then be calculated by solving the three-fold integral (see e.g. \cite{dwek:2005a}):
\small
\setlength{\mathindent}{0pt}
  \begin{equation}
    \label{eq:tau1}
      \tau(E_{\gamma},\,z) = \int_{0}^{z} {\rm d}\ell ({\rm z'}) \, \int_{-1}^{1} d\mu\,\frac{1-\mu}{2}  
                        \int_{\epsilon'_{th}}^{\infty} d\epsilon'\,n(\epsilon',z') \,\sigma_{\gamma\gamma}(\epsilon',E'_\gamma,\mu) %\, \nonumber \\
  \end{equation}
\normalsize
\setlength{\mathindent}{2pt}
where $\mu  =  \cos{\theta}$,  
$n\left(\epsilon\,\right)$  is EBL photon number density and 
${\rm d}\ell({\rm z})$ the distance element
(expressed in co-moving coordinates).
 
 %------------------
% FIGURE: Fazio Stecker plot F3
\begin{figure}[tbp]
\centering
\includegraphics[width=0.5\textwidth]{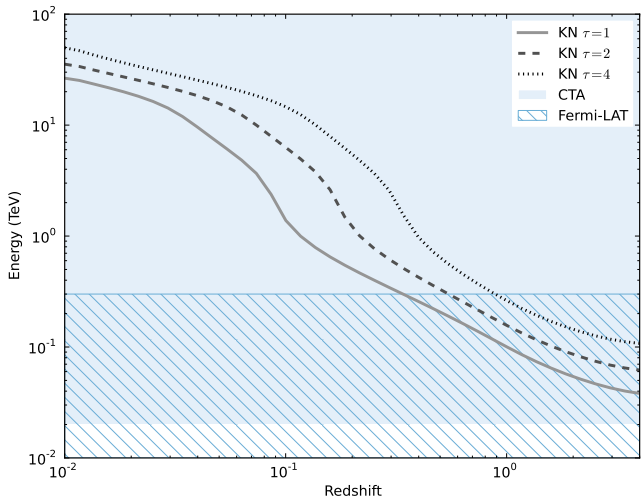}
\caption{Energy at which an optical depth of $\tau=$1, 2, and 4 is reached vs. redshift for the EBL model from \cite{kneiske:2010a} (KN, minimal EBL). Also shown are approximate energy ranges of CTA (20\,GeV-100\,TeV; shaded) and \FermiLAT (100\,MeV-300\,GeV; hatched).}
\label{Fig:FazioSteckerPlot}
\end{figure}
 %------------------

Fig.~\ref{Fig:EBLAttenuation} displays the resulting attenuation due to the EBL for VHE $\gamma$-ray sources at different redshifts and for different EBL models (see Fig.~\ref{Fig:EBLModels}). Several features can be identified:
\begin{itemize}
\item At low energies ($<80$\,GeV) the spectrum is practically unattenuated (with the exception of extreme EBL models at high redshifts, i.e. the fast evolution model from Stecker et al. at $z=2$). Since this energy range will be sampled by the CTA with high 
statistical precision, it will be possible to measure the unabsorbed spectrum.
\item At energies from 80\,GeV to 2\,TeV the strength of the attenuation is increasing due to the EBL photons in the optical to near-infrared range peak of the EBL density.
\item At energies between 2 to 10\,TeV a flattening of the attenuation is expected at low redshifts 
(as seen for $z=0.1$ in Fig.~\ref{Fig:EBLAttenuation} for FR and GEN models), 
due to the $\sim \lambda^{-1}$ behavior of the EBL density in the near to mid-infrared, resulting in a constant attenuation. Such a modulation of the EBL attenuation has been considered as a possible key signature for EBL attenuation
\cite[e.g.][]{guy:2000a,aharonian:2002a,costamante:2003a,orr:2011a}.
Unfortunately, the intrinsic weakness of the sources combined with the sensitivity of the instruments makes it difficult to probe such a feature with previous or current generation experiments
(but see \cite{orr:2011b}).
\item At energies around 10\,TeV the turnover in the EBL density towards the far-infrared peak of the EBL results in a strong attenuation, effectively resulting in a cut-off in the measured spectra.
\end{itemize}
The strength and the position of these features vary with the distance of the VHE sources, the assumed EBL model, and the overall EBL density. This can also be seen in Fig.~\ref{Fig:FazioSteckerPlot}, 
which shows the energies where the EBL gets optically thick for different redshifts (i.e.\ the optical depth reaches unity or higher).

It should be noted that only distant extragalactic VHE \g-ray emitters suffer from the absorption by the EBL. For galactic sources the attenuation effect from the EBL is negligible up to energies of about 100\,TeV. For higher energies, the absorption by the photon field of the CMB starts to be important. Galactic diffuse photon fields also produce a weak attenuation signature, which might be seen with CTA \cite{moskalenko:2006a}.

The produced e$^+$/e$^-$ pairs can initiate electromagnetic
cascades by upscattering low-energy CMB photons to high
energies; these latter photons would produce, yet again,
e$^+$/e$^-$ pairs by interacting with EBL photons (see, e.g., \cite{aharonian:1994a,neronov:2009a}).
The $\gamma$-rays resulting from the cascade would mostly appear at lower energies than the ones  accessible by CTA but they would modify the overall energy spectra of the sources, which is important for their modeling.
If the strength of intergalactic magnetic fields and interaction length are low enough, the cascade component would appear to arrive from the direction of the source. Higher magnetic fields would isotropize the cascade component.

Since the EBL attenuation alters the VHE $\gamma$-ray spectra of all distant sources, understanding the EBL's properties are crucial for studying the intrinsic source physics. On the other hand, VHE $\gamma$-rays from distant sources can be used as probes for the intervening photon fields. This requires some assumptions about the intrinsic spectra, which, e.g., can be derived from nearby sources or unattenuated parts of the spectrum. In addition, EBL attenuation should be a global phenomenon affecting the spectra of all sources in a consistent way. In particular, sources at the same distance should show the same attenuation imprint from the EBL.

%--------------------------------------------------------------------------------
% The CTA EBL \& cosmology physics case
%--------------------------------------------------------------------------------
\section{The CTA EBL \& cosmology physics case}

%--------------------------------------------------------------------------------
\subsection{Disentangling intrinsic blazar spectra from EBL effects: precision measurement of the EBL density} 

%- Fundamental difficulty to disentangle

%- Solution is to simultaneously measure absorbed and unobsorbed spectra
%- Also spectra in different flux states and of different sources: EBL imprint stays always the same
% - Mention better modeling

% The disentengling will lead to a precision measurement of integrated EBL.
% Here is to mention that hard sources up to 1-10 TeV are the best ones to constrain
% opt-midIR. must not be distant (<0.3)
% Distant sources (z>0.2) with spectra up to 300-500 GeV needed to measure EBL in UV.
% GRBs may help (see also Fermi paper)
% need EHBL (photons up to 30 TeV) to probe FIR

%- Fundamental difficulty to disentangle
There is a fundamental difficulty to distinguish between intrinsic blazar
effects (e.g.\  intrinsic cut-offs or possible multi-zone emission regions) and
the absorption effect caused by the EBL. This ambiguity is also the main reason
why, with already more than 40 extragalactic sources measured by H.E.S.S.,
MAGIC and VERITAS, only upper limits on EBL density have been derived yet (as in, e.g., \cite{aharonian:2006:hess:ebl:nature}).  The
number of  sources detected at VHE will increase further in the next 5 years,
in particular owing to the upgrades of the existing facilities aiming at enhanced sensitivity and lower the energy thresholds: MAGIC\,II, H.E.S.S.\,II
and VERITAS\,II. However, the real boost in the field will be achieved with CTA only,
when it will become possible to measure simultaneously and with high precision
both the unattenuated (usually at energies below 100 GeV) and attenuated (at 
higher energies) parts of the AGN spectra.
Such measurements will allow to disentangle between intrinsic features, which 
can vary from source to source and even for the same source depending on its 
flux state, and the EBL induced signature, which solely depends on the redshift of 
the source. Also, precise measurements at energies $>$ 10-20 GeV
(the foreseen energy threshold of CTA) in combination of multiwavelength data 
would ease and improve the modeling accuracy of AGNs \cite{mankuzhiyil:2010a}.

% The disentengling will lead to a precision measurement of integrated EBL.
Disentangling the intrinsic AGN features and the EBL induced
attenuation signature in the measured AGN spectra will unambiguously lead to a precise
measurement of the EBL density at a redshift of $z=0$.  The precision of the measurement
will depend on the richness of the AGN sample available, in particular on the
number of sources at different redshifts to enable a robust modeling of the
intrinsic features of the sources.
The power of CTA to simultaneously measure the intrinsic and the EBL attenuated part of an energy spectrum is investigated in Section~\ref{Sec:DisentaglingIntrFromEBL} utilizing the measured VHE of  the blazar PKS\,2155-304 as a template.

In individual sources an intrinsic break or cut-off cannot be excluded.
However, with CTA it is estimated (see \cite{inoue:2010a}) 
that over a period of 5 years conservatively 100 new AGN
sources could be expected to be discovered. This gives a sizable source sample
to check for global features/trends in the spectrum as a function of the
expected opacity (see, e.g., \cite{meyer:2012a} for a
study on the current VHE AGN sample), or redshift (see, e.g., \cite{behera:2008a}
for a simulated AGN sample) allowing
to exclude biases due to source specific features. Combined with statistical
studies of a large number of objects, precision measurements point to a unique
and powerful method to probe the EBL that is difficult to achieve with
present-day instruments.  

%Best sources to measure EBL
Owing to the dependence of the pair-production cross section on the energy of
the participating photons, distinct AGN types are crucially important in the
indirect EBL measurements at different wavelengths.  For measuring the EBL in
UV, the sources must be distant to experience a measurable absorption in UV and
they must be sufficiently bright to provide enough flux for a CTA measurement.
Moreover, their intrinsic spectra must reach energies up to 300--500 GeV in order to
connect from UV to the optical regime.  Therefore, distant bright AGNs, mostly
FSRQs (e.g.  3C\,279 \cite{albert:2008:magic:3c279:science} and PKS\,1222
\cite{aleksic:2011:magic:pks1222}), and maybe GRBs if extending to high enough
energies are the sources with the best potential to probe the EBL in UV.  For
measuring the EBL in optical to mid IR regimes, AGNs with hard spectra
extending to few TeV are the ones providing the best results. The redshift of
the sources must not be too large (z$<$0.5) in order to avoid a total
absorption of the flux at energies above 1 TeV.  This is a guaranteed science
case, since good candidates with hard energy spectra already exist:
1ES\,0229+200 \cite{aharonian:2007:hess:0229}, RGB J0710+591 \cite{acciari:2010a},
1ES\,1101-232 \cite{aharonian:2007:hess:1es1101}.  A similar case can be made
for the far IR regime. There, few close by sources (e.g. Mrk 421
\cite{blazejowski:2005a}, Mrk 501 \cite{aharonian:1997b}, M\,87
\cite{aharonian:2006:hess:m87:science}, IC\,310 \cite{aleksic:2010:magic:ic310}) are the best ones
provided they produce intrinsic spectra up to energies of at least E$>$20\,TeV.

%Will set constraints on exotic contributions
The promising technique of the indirect EBL measurements 
can be combined with measurements from deep source counts at optical to IR wavelengths.
Through this, the contribution to the EBL from faint unresolved sources or any exotic populations 
can be disentangled from relatively bright and known populations.

%--------------------------------------------------------------------------------
\subsection{Star formation history and the early universe} 
The cosmic star formation rate density (SFRD), one of the fundamental
quantities of cosmology,  strongly evolves over the redshift range $z=0$ to
$z=2$ with a peak expected at redshifts between 1 and 2.  While the SFRD up to
$z<1$ is reasonable well measured from different tracers, at higher redshifts
the uncertainties are large. Since stellar and dust emissions are expected to
be the dominant contributors to the EBL, a precise measurement of the EBL
density can provide strong constraints on the SFRD \cite{raue:2012a}.
The science output is guaranteed since the SFRD is a highly debated topic of
the modern cosmology. It is estimated that the precision of the SFRD
determination, which can be achieved using VHE spectra, will be on a level of
20--30\% (based on \cite{blanch:2005a}).

%--------------------------------------------------------------------------------
The end of the dark ages of the universe - the epoch of reionization - is a topic
of great interest \cite{barkana:2001a,ciardi:2005a}. Not (yet) accessible via direct observations, most of our
knowledge comes from simulations and from integral observables like, e.g., the
cosmic microwave background. The first (Population III) and second (Population II) generations
of stars are natural candidates for the sources of reionization. If the first
stars are hot and massive, as predicted by simulations \cite{bromm:2004a,glover:2005a}, their UV photons
emitted at $z > 5$ would be redshifted to the near-infrared regime and could leave
a unique signature on the EBL density \cite{santos:2002a}. If the EBL contribution from lower
redshift galaxies is sufficiently well known (e.g., as derived from source
counts) upper limits on the EBL density can be used to probe the properties of
early stars and galaxies \cite{raue:2009a}. Combining detailed model calculations with redshift
dependent EBL density measurements could enable to probe the
reionization history of our universe. Detection of spectral cutoffs
in sources at very high redshifts ($z>5$) would directly probe the evolving UV
background at these redshifts, providing invaluable insight into the cosmic
reionization epoch.

While blazars may be difficult to detect much beyond $z\approx2$,
this may be feasible for GRBs.
GRBs are the most luminous and distant gamma-ray sources known in the Universe,
typically arising at redshifts $z \sim 1-4$,
which correspond to the peak epoch of cosmic star formation activity.
Furthermore, they are known to occur at least up to $z \sim 8-9$ \cite{tanvir:2009a,salvaterra:2009a},
well into the cosmic reionization era,
and possibly even beyond, out to the very first epochs of star formation in the Universe.
The recent detections by {\it Fermi}/LAT of dozens of GRBs
including GRB 080916C at $z = 4.35$ \cite{abdo:2009:fermi:grb080916C}
clearly demonstrate that at least some GRBs have luminous emission extending to few tens of GeV \cite{oh:2001a,inoue:2010b}.
Moreover, the duration of the multi-GeV emission is seen to last up to several thousand seconds,
in agreement with reasonable theoretical expectations for the high-energy afterglow emission see e.g.\ \cite{inoue:2010b}.
Thanks to CTA's much greater effective area compared to {\it Fermi}/LAT (by about a factor of $10^{4}$ at 30 GeV
if using CTA array layout $B$)
together with its rapid slewing capability (180 degrees within 20 seconds),
there are good prospects for high precision measurements of the GeV-TeV spectra of GRBs at $z > 1$,
as well as the detection of GRBs at $z \gtrsim 4$ and beyond, a regime that cannot be explored with AGNs.

%--------------------------------------------------------------------------------
\subsection{Particle physics and the unknown}
Data collected so far by H.E.S.S., MAGIC 
and VERITAS on absorption for AGNs have
constrained the maximum EBL density to a level close to the lower limits from integrated galaxy
counts. Once CTA goes online and significantly improves the indirect EBL measurements, several scenarios are possible.
In the first scenario the joint efforts
of direct and indirect measurements converge and the EBL density is resolved.
If the EBL density is well measured,
EBL absorption in VHE spectra can be used as a distance
indicator, somewhat analogues to what is done for Supernovae of type 1A but
with very different systematic uncertainties (see \cite{blanch:2005a}). The
evolution of such an attenuation signature with redshift would allow us
to put constraints on cosmological models and measure the Hubble parameter and
cosmological densities. 
The proposed method would provide
independent and complementary constraints
on the cosmological parameters with different systematic uncertainties
to what projects as LSST \cite{ivezic:2008a:lsst} or PLANCK \cite{ade:2011a:planckmission} are planning to reach.
An independent distance measurement of AGNs using the
EBL absorption feature in their VHE spectra is also
important for getting redshifts for many BL Lacs, since only 50\% 
of all BL Lacs have reliable redshift measurements.

In a second scenario, the EBL level determined from the VHE observations
will be remain to be a factor of two to three higher than the level 
resolved through the galaxy counts. 
This would point to a large population of yet hidden sources,
maybe also to a truly diffuse background such as due to decaying big-bang relict particles. 
In case the discrepancy between the lower and upper limits will be restricted to a narrow wavelength range,
the search for the missing background photons will be easier.

In an alternative scenario, the upper limits on EBL from indirect measurements 
will contradict lower limits from galaxy counts. This opens up a completely new 
window in particle and astroparticle physics since one of the two following options 
must be true:
either the intrinsic source physics is fundamentally different than the models predict
or the properties of gamma ray propagation through space are fundamentally different
than we know today, which can become a major challenge for the
modern physics (see e.g. \cite{deangelis:2007a,barrio:2011a}).

%--------------------------------------------------------------------------------
% Investigating the CTA response
%--------------------------------------------------------------------------------
\section{Investigating the CTA response}

%--------------------------------------------------------------------------------
\subsection{Simulations}

%------------------
% FIGURE: LowEnergiesPKS2155PLResults F7
\begin{figure*}[tbp]
\centering
\includegraphics[width=0.45\textwidth]{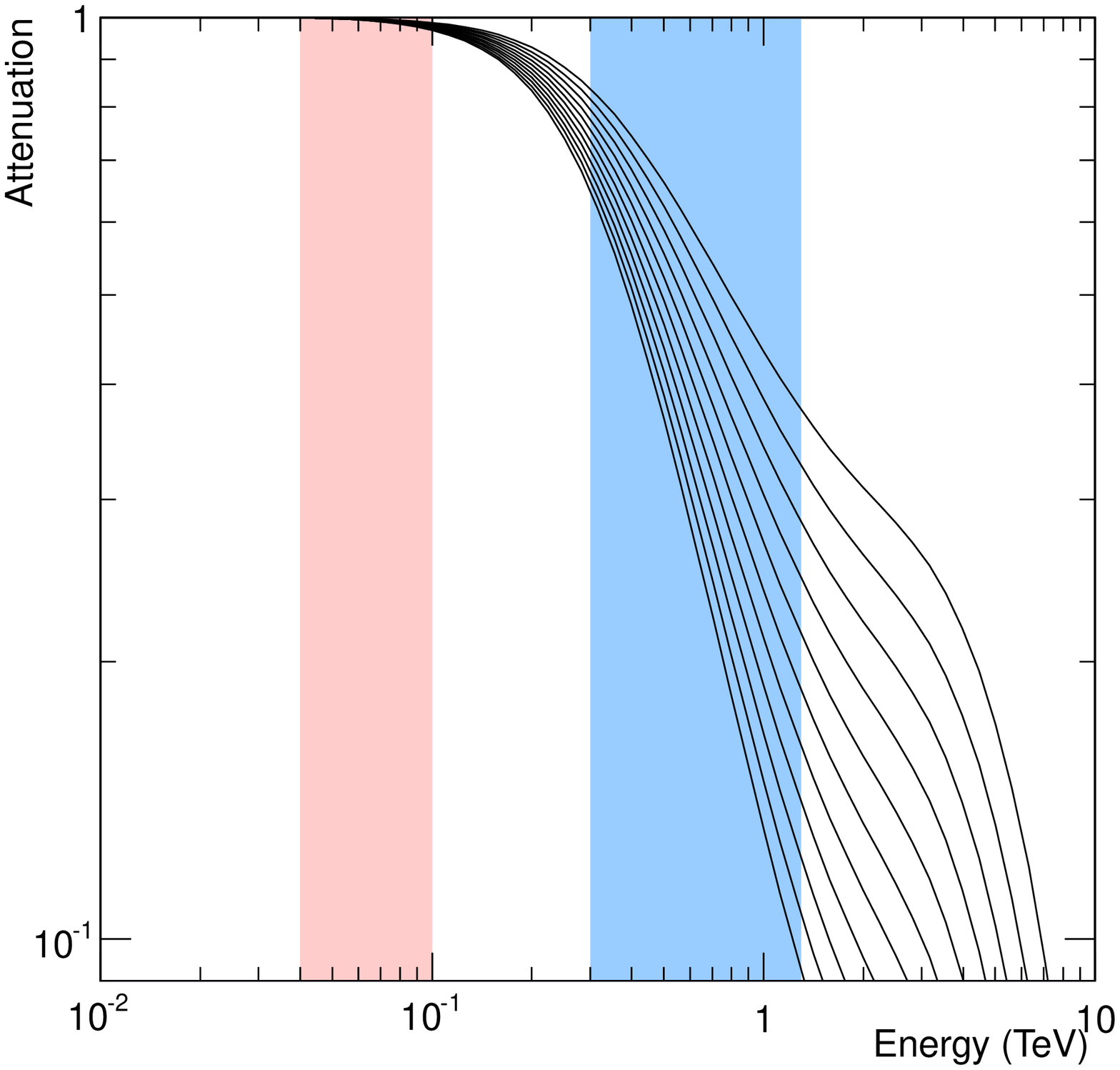}%\hfill%
\includegraphics[width=0.45\textwidth]{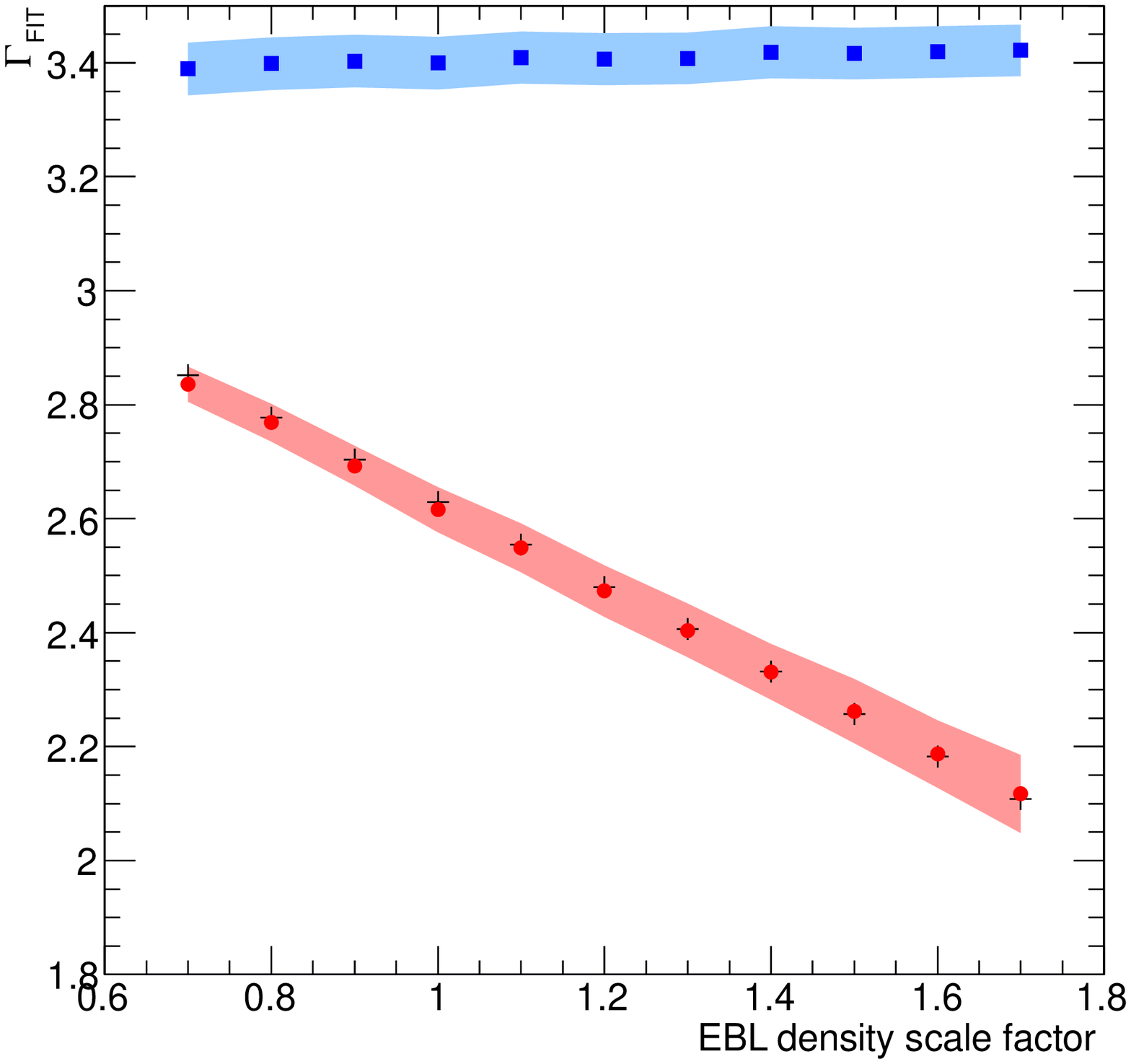}
\caption{
Results from the power law fit to the simulated spectrum in the low and high
energy band using the PKS\,2155-304 quiescence spectrum, the EBL model from
\cite{franceschini:2008a}, and CTA array layout $B$ (20\,h exposure).
\textit{Left:} Attenuation of the VHE 
$\gamma$-ray flux resulting from the scaled EBL models (lines). 
The colored boxes in the background mark the fit range for the power law at low (red) and high
(blue) energies. \textit{Right:} Spectral index $\Gamma$ from the fit of a
power law to the low (red) and high (blue) energy range versus the scaling
factor of the EBL model. The color shaded bands denote the error on the
spectral index from the fit (RMS of the spectral index distribution). Black
crosses mark the intrinsic spectral index that has been utilized.
For discussion see main text.}
\label{Fig:LowEnergiesPKS2155PLResults1} 
\end{figure*}
 %------------------

For the studies presented here the CTA physics response simulation tools presented in \cite{hinton:2011a} are used.
In short, a source intrinsic spectrum is assumed in each case and then folded with a given CTA response. The source is located at the center of the telescopes' field of view and the background (especially important at low energies) is assumed to be measured with a 5-times better statistics than the exposure of the source region. Only significant, i.e.\ larger than 3 standard deviations, reconstructed flux points have been considered in the analysis. Furthermore, the systematic uncertainty is considered to be 1\% of the background level, and we require the signal to be at least 3 times higher than that. The performance files for 20$\deg$ zenith angle (fixed) are used and no correction for source motion along the sky is made.

%------------------
% FIGURE: LowEnergiesExample F4
\begin{figure}[tbp]
\centering
\includegraphics[width=0.5\textwidth]{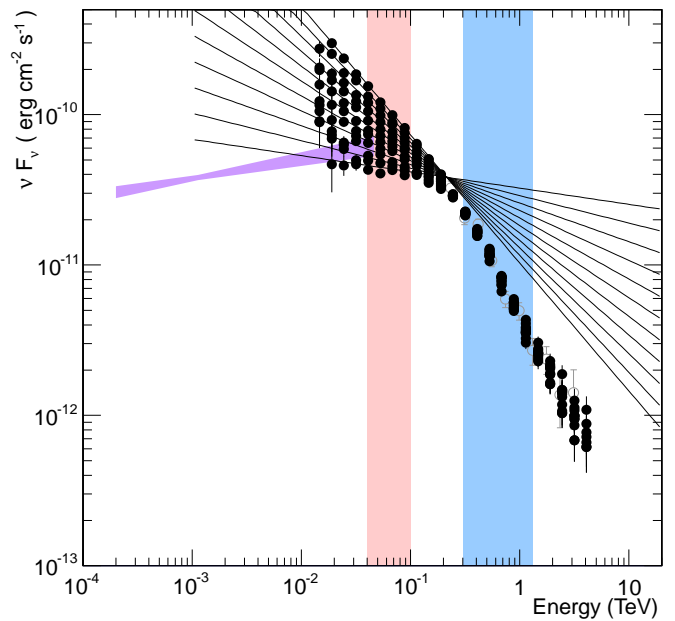}
\caption{Example of simulated spectra for different EBL densities. The base
spectrum assumed is the quiescence state spectrum of PKS\,2155-304 ($z=0.116$),
EBL model is from \cite{franceschini:2008a}, used with different scalings. Shown are: the measured spectrum
(grey markers), the simulated spectra for different level of the EBL density
(black markers) and the corresponding assumed intrinsic spectra (black lines),
the source spectrum in the GeV energy range as measured by \FermiLAT
\cite{abdo:2009a:fermi:brightsourcelist:agn} (not simultaneous, purple butterfly), and the energy
ranges, which are used to determine the slope of the simulated spectrum at low
(light red) and high (blue) energies. For details on the data and methods see \cite{raue:2010a}.} 
\label{Fig:LowEnergiesExample} 
\end{figure}
 %------------------

An example of such a simulated spectrum is shown in Fig.~\ref{Fig:LowEnergiesExample}, where the measured spectrum of the blazar PKS\,2155-304 is used together with different assumptions about the EBL density (here a scaled version of the EBL model by \cite{franceschini:2008a}) to calculate the spectrum as it would have been detected by CTA in a 20\,h exposure (CTA array layout: B) (more details can be found in \cite{raue:2010a}).

%--------------------------------------------------------------------------------

\subsection{Results}

\subsubsection{Disentangling intrinsic blazar spectra and EBL effects.}
\label{Sec:DisentaglingIntrFromEBL}

One of the main features of CTA will be a high sensitivity in the energy range between 20 and 100\,GeV, an energy range which holds the possibility to directly sample parts of the energy spectrum of a source, which are not affected by the EBL attenuation. With the difference between the measured spectrum in the unabsorbed part of the VHE spectrum and in the absorbed part of the spectrum, especially if studied for several sources with good statistics, 
the strength of the near-IR EBL can be derived.
Here, exemplarily, the case of  PKS\,2155-304 is studied following the approach described in \cite{raue:2010a}.

%------------------
% FIGURE: LowEnergiesPKS2155PLResults II
\begin{figure*}[tbp]
\centering
\includegraphics[width=0.8\textwidth]{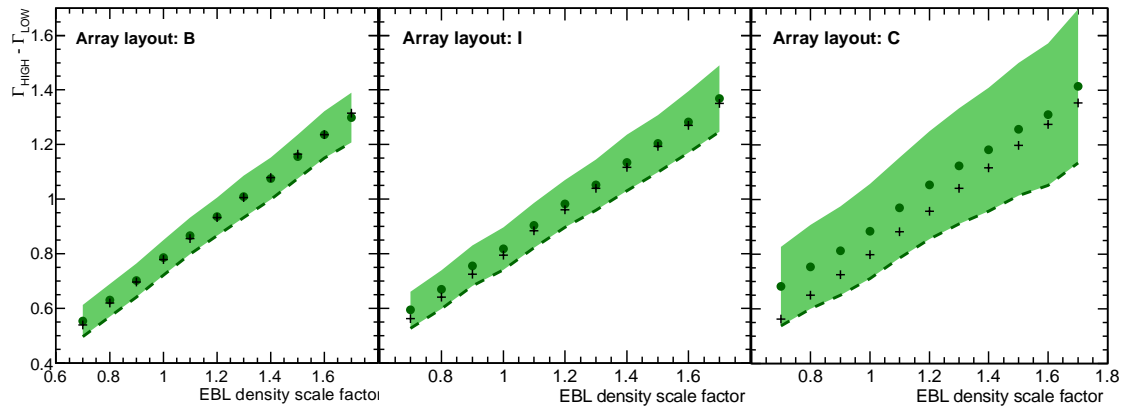}
\caption{Spectral break between the power law fits in the low and high energy bands for different CTA array layouts (spectrum: PKS\,2155-304 quiescence; EBL model: \cite{franceschini:2008a}; exposure: 20\,h).
Circles mark the reconstructed spectral break derived from the simulated data while crosses mark the spectral break between the simulated spectral index and the reconstructed index in the high energy band.
The error bands are derived from error propagation.
For more details on the anlysis see \cite{raue:2010a}.
}
\label{Fig:LowEnergiesPKS2155PLResults2} 
\end{figure*}
 %------------------
 
For different levels of the EBL density, generated by scaling the EBL model from \cite{franceschini:2008a}, the intrinsic source spectrum is reconstructed and
used to simulate the corresponding spectrum as measured by CTA (Fig.~\ref{Fig:LowEnergiesExample}). The resulting attenuations for the scaled EBL models are shown in Fig.~\ref{Fig:LowEnergiesPKS2155PLResults1} left panel.
In the CTA simulated spectrum two distinct regions in energy can be identified: (i) the unabsorbed region, roughly from 40 to 100\,GeV for the redshift of PKS\,2155-304 of $z=0.116$ (red band), where no or little EBL attenuation is present and the intrinsic spectrum can be observed directly, and the (ii) absorbed region at energies $>100$\,GeV, where the EBL attenuation can be measured (blue band). In each of the two bands the spectrum is characterized by a power law $dN/dE \sim E^{-\Gamma}$ and the derived photon indices $\Gamma$, shown in Fig.~\ref{Fig:LowEnergiesPKS2155PLResults1} right panel, are then investigated. The fit in the low energy band reproduces very well the assumed intrinsic spectral indices (red dots vs black crosses in  Fig.~\ref{Fig:LowEnergiesPKS2155PLResults1} right panel).

Fig.~\ref{Fig:LowEnergiesPKS2155PLResults2} displays the strength of the spectral break between the two power laws $\Gamma_{LOW} - \Gamma_{HIGH}$ for different CTA array layouts, with the error bands calculated via error propagation. For array layout $B$ (best low energy performance) an EBL density 
a step in scale factor of 0.1 corresponds to one to two standard deviations difference in the strength of the break. Array layout I provides a comparable but slightly worse performance while the decreased low energy sensitivity of array layout $C$ makes studies of this type difficult (resolution factor two to three worse and the intrinsic index cannot be reconstructed).
Note that a scaling of 0.1 corresponds to an EBL density of $\sim$1\,nW\,m$^{-2}$sr$^{-1}$ at 2\,$\mu$m which is of the same order as the error on the lower limits from integrated source counts at this wavelength.
Since EBL attenuation is a global, redshift dependent phenomenon, in general, a combined fit to all available VHE data should be used to derive limits on the EBL density \citep[see e.g.][]{mazin:2007a}.

While for many of the up to now detected extragalactic VHE sources the energy spectrum in the VHE range is well described by the extrapolation of the spectrum in the HE range folded with an EBL attenuation some of the sources show an intrinsic break between the two energy regimes (e.g. PKS\,2155-304) \cite{abdo:2009:fermi:tevselectedagn}. For individual sources, such an intrinsic break could lead to a wrongly determined EBL density when blindly applying the methods outlined above. The solution, again, lies in utilizing a sample of source: while intrinsic features are different between the sources the EBL attenuation signature can be precisely determined for each of them enabling one to disentangle the two.

\subsubsection{Distant sources, GRBs}
\label{Sec:grbssection}

Due to the low energy threshold of CTA and its high sensitivity CTA will have the opportunity to detect even distant GRBs. 
Whereas a discussion on GRB physics is beyond the scope of this article, GRBs start playing a crucial
role in understanding EBL density at high redshifts \cite{abdo:2010:fermi:eblconstrains}. A common expectation is
that only GRBs will provide high enough $\gamma$-ray luminosity to detect a  source located at high redshifts ($z>2$). 
Also AGN in extreme flaring states could be detected at $z>2$ (see Figure~6 in AGN paper, same issue).
As mentioned before, $\gamma$-ray sources beyond a redshift of two can be used as beacons
probing the star formation rate density and, in particular, its peak position.

\begin{figure*}[tbp]
\centering
\includegraphics[width=0.45\textwidth]{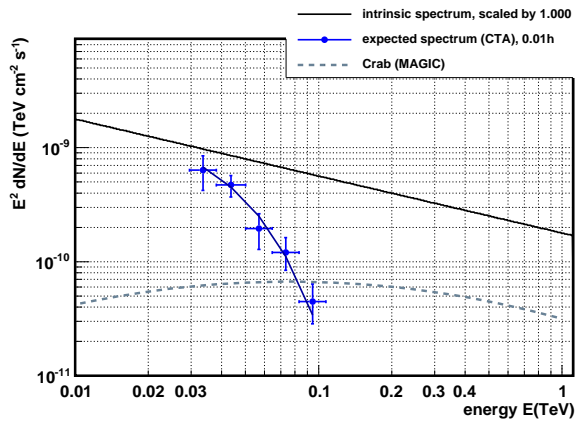} %\hfill
\includegraphics[width=0.45\textwidth]{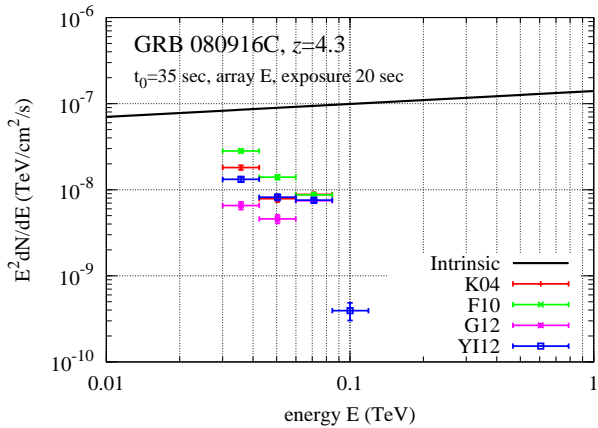}
\caption{
Simulated energy spectrum of GRB 080916C ($z=4.3$) if measured with CTA. %\newline
{\it Left panel:}
Using the CTA array layout $B$ (low energy threshold)
and assuming EBL from \cite{dominguez:2011a}.
The intrinsic spectrum is assumed to follow  $2.9\times 10^{-9} \times
(E/\mathrm{TeV})^{-2.16} [\mathrm{TeV}^{-1} \mathrm{cm}^{-2} \mathrm{s}^{-1}]$
and the duration of the measurement is 45\,s (T$_0$=55--100s, interval "e") as measured with $Fermi$/LAT \cite{abdo:2009:fermi:grb080916C}. 
A clear detection can be made and spectral shape can be measured from 30 to 100\,GeV.  The result is
basically identical if using CTA array layouts $E$ or $I$. For the array layout $C$ (high energy
threshold), no detection of this GRB can be made. %\newline
{\it Right panel:}
Using the CTA array layout $E$ and exposure time 20 sec for the interval "d" \cite{abdo:2009:fermi:grb080916C}, 
i.e. the assumed source flux is $dN/dE=1.4 \times 10^{-7} (E/{\rm TeV})^{-1.85} {\rm cm^{-2} s^{-1} TeV^{-1}}$.
Here we compare the EBL models of Kneiske et al.~``best fit'' \cite{kneiske:2004a} (red), 
Finke et al.~\cite{finke:2010a} (green), Gilmore et al.~\cite{gilmore:2012b} (magenta), and Inoue et al.~(private communication, blue).
%Here we compare the EBL models of Y.~Inoue et al. \cite{inoue:2011a} (blue) and Kneiske et al. ``best fit'' \cite{kneiske:2004a} (red). 
%Due to the lower absorption in case  of the EBL model of Y.~Inoue et al. 
%the spectral shape of the GRB emission can be measured much better.
For the array layout $C$, again, no detection of this GRB can be made. 
}
\label{Fig:GRB080916C_45sec} 
\end{figure*}

The power of CTA to detect distant GRBs can be demonstrated by utilizing the
\FermiLAT detection of the extraordinary GRB\,080916C, which took place at a
redshift of $z=4.3$ \cite{abdo:2009:fermi:grb080916C}. 
In Fig.~\ref{Fig:GRB080916C_45sec} we compare simulated spectra 
as they were measured by CTA of the GRB.
In the left panel, we assume the instrinsic spectrum as it was measured 
by \FermiLAT for the time-span $T_0=55-100$\,s (45\,s, interval "e" as defined in \cite{abdo:2009:fermi:grb080916C}).
The result is that a clear detection can be made, here on an example of CTA array layout $B$ 
(see \cite{bernloehr:2012a:ctamcpaper} for details on individual array configurations. 
For this array layout, it would also have been possible to derive a detailed
spectrum. Comparing performance of different array layouts for this physics case, we see only marginal improvement using the array 
layout $B$ (low threshold) instead of $E$ or $I$ (balanced arrays): 
the resulting error bars are only few percent better and the reconstructed energy range is identical.
However, for the array layout $C$ (high threshold), no detection of this GRB can be achieved and, consequently, no spectrum can be derived.  
Since GRBs are short lived events, a high repositioning speed of CTA
is required.
In the right panel of  Fig.~\ref{Fig:GRB080916C_45sec} we see a simulation of the GRB080916C spectrum
for the interval "d" \cite{abdo:2009:fermi:grb080916C})\footnote{
The assumed source flux is $dN/dE=1.4 \times 10^{-7} (E/{\rm TeV})^{-1.85} {\rm cm^{-2} s^{-1} TeV^{-1}}$,
whereas the index -1.85 is derived from a fit to the LAT-only data (Figure S4 in the Supplement to
\cite{abdo:2009:fermi:grb080916C} and private communication from the Fermi-LAT team).}. 
The intrinsic spectrum is harder and the flux level
is higher than for the interval "e", which results in clear detection even for an integration time of 20s.
Results for four different EBL models 
Kneiske et al.~``best fit'' \cite{kneiske:2004a} (red),
Finke et al.~\cite{finke:2010a} (green), Gilmore et al.~\cite{gilmore:2012b} (magenta), and Inoue et al.~(private communication, blue)
are presented,
%Results for two different EBL models Y.~Inoue et al. \cite{inoue:2011a} (blue) and Kneiske et al. ``best fit'' \cite{kneiske:2004a} (red)
%are presented, 
which show clear difference and permits distinguishing between the models.
We note that catching a GRB at a time $T_0<55$\,s after the onset of the GRB, which would correspond
to the interval "d" in case of GRB080916C, is challenging even if CTA would be able to reposition to any pace in the sky within 20\,s.
It is, however, withing the reach of the instrument. Observations at $T_0>55$, corresponding to 
the interval "e" in case of GRB080916C, are easier to achieve and can be considered as a secured case.
It is important to mention that GRB080916C was the brightest GRB observed by the Fermi/LAT so far.
The probability that GRBs with sufficiently high flux will be observed by CTA within its life time 
has been addressed by several authors \cite{kukawa:2011a,gilmore:2012a} finding that CTA, if having a low energy threshold of 20--30GeV
would be able to detect  0.1--0.2 GRBs per year during the prompt phase and  about 1 per year in the afterglow phase.

As discussed by a number of authors \cite{oh:2001a,inoue:2010b,inoue:2011a},
UV radiation with sufficient intensities to cause the reionization of the intergalactic medium (IGM)
are also likely to induce appreciable gamma-ray absorption in sources at $z \gtrsim 6$
at observed energies in the multi-GeV range,
with a potentially important contribution from Pop III stars.
Measurements of these effects can thus provide
important cross-checks of current models of cosmic reionization,
a unique and invaluable probe of the evolving UV EBL
during the era of early star formation,
as well as a test for the existence of the yet hypothetical Pop III stars.
For more details we refer to the GRB paper in this issue.

\subsubsection{Precision measurement of today's EBL}\label{Sec:PrecisionMeasurementOfZ0EBL}

The measured energy spectra of AGNs in the energy range between 100\,GeV and
few TeV follow usually a smooth shape. For most of the measured sources, a
simple power law fit is sufficient to describe the available data well, whereas
for sources in a flare state (like the flare of PKS\,2155-304 in 2006) or with
a generally high emission state (like Mkn\,421), either a curved power law  or
a power law with a cut-off are successfully used. The curved power law (also
known in the literature as the double-log parabola) is expected to describe the
spectra well at energies close to the position of the Inverse-Compton peak. The
power law with a cut-off instead is the expected behavior of a source  which
does not provide necessary conditions for acceleration of charged particles to
sufficiently high energies. All scenarios, assuming the one-zone model,
which seems adequate to describe most VHE spectra of AGNs rather well, do have one common feature: the
measured spectra can be described by smooth functions, i.e., no features,
wiggles or pile-ups are expected, especially after de-convolving the spectra
for the effect of the EBL absorption.  This property can, therefore, be used to
distinguish between different overall EBL levels in the optical to infrared
regime: whereas the "correct" EBL model and level will produce a smooth
intrinsic AGN spectrum, an "incorrect" EBL level would result in a signature
(in form of well defined wiggles) in the reconstructed intrinsic spectrum 
\cite{dwek:2005b,raue:2010a}.
Possible spectral features in the intrinsic spectra could complicate the
analysis, though other methods like investigating the spectral properties of
variable sources on short time-scales or using multiwavelength based modeling might help to
disentangle source intrinsic and EBL effects (see Sec 2.1). 
It is also important to observe the same effect on spectra of different sources to become
independent of possible intrinsic source features.
Another caveat is that the form of the wiggles will have some dependency on the exact
spectral shape of the "correct" EBL model, which is not known a-priori.
Though the state-of-the-art EBL models agree within 10\% on the expected EBL shape
(with the only exception being the model of \cite{stecker:2006a}),
a possibility of a significantly different shape, e.g., due to a contribution from early stars,
cannot be excluded.

The strength of the method is that the EBL signatures in the reconstructed AGN
spectra will not only be visible (measurable) in the case where the assumed EBL
level is higher than the real one, but also in case the assumed EBL level is
lower than the real one. It is, therefore, the first indirect method to really
measure the EBL density at z=0.

%------------------
% FIGURE: Mid energies - example spectrum F10
\begin{figure*}[t]
\centering
\includegraphics[width=0.65\textwidth]{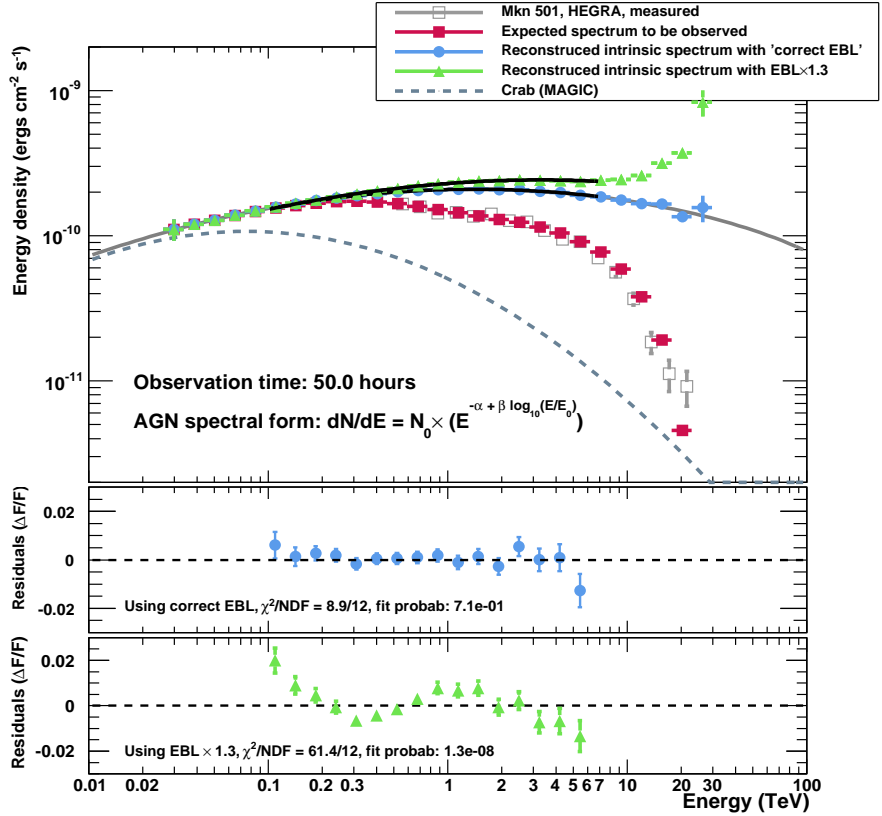}
\caption{
Search for signatures at mid-VHE in the Mkn\,501 spectrum.
\textit{Top panel:} The Mkn\,501 spectral energy distribution (SED) at very high
energies. Shown are: measured spectrum by HEGRA (grey open squares),
simulated CTA spectrum assuming the array layout $I$ (red filled squares),
assumed intrinsic spectrum of Mkn\,501 (grey solid line),
reconstructed intrinsic Mkn\,501 spectrum (blue filled circles) and reconstructed
intrinsic spectrum of Mkn\,501 assuming an EBL scaling factor of 1.3 (green triangles).
The Crab Nebula spectrum (grey dashed line) is shown for comparison.
\textit{Middle panel:} Residuals between the best fit to the SED in case of CTA measurement and the spectral points from the intrinsic spectrum using the correct EBL density.
\textit{Bottom panel:} Residuals between the best fit to the SED for a CTA  measurement and the spectral points from the intrinsic spectrum reconstructed using the scaled EBL model.
A clear and significant signature (wiggles) in the residuals is visible, which is quantified by
a low probability of the fit.}
\label{Fig:MidEnergyExampleOfMethod}
\end{figure*}
 %------------------
 
\paragraph{Simulation example}
The method is illustrated in
Fig.~\ref{Fig:MidEnergyExampleOfMethod} for the VHE spectrum of Mkn\,501. The assumed
spectral shape and the flux level of the intrinsic spectrum are adapted to the
average flux measured by HEGRA  \cite{aharonian:1999b} during the outburst of the source in 1997 (110\,h observation time): the original data are shown by grey open squares in the upper panel of the figure.
The simulated CTA spectrum calculated using the 'correct EBL' (model from \cite{raue:2008b})
is shown in red, whereas the assumed intrinsic spectrum of Mkn\,501
is shown by the solid grey line and the reconstructed intrinsic spectrum
is shown by the blue filled circles.
The effect of the mis-reconstruction of the intrinsic spectrum is shown for the example of
an EBL scaled by a factor of 1.3: the reconstructed intrinsic spectrum (green filled triangles)
clearly shows wiggles in the fit range. The effect of the wiggles is more visible in the lower panel
of the figure where the residuals to the best fit function are shown.
The wiggles are quantified by a fit in the energy range
between 100\,GeV and 7\,TeV, well before a possible pile-up in the
spectrum arises,
which can be additionally used to rule out an EBL realization \cite{dwek:2005a,mazin:2007a}.
The choice of the fit range is made in order not to bias the
result by the level of the EBL above 10\,$\mu$m, to which  the VHE spectra are
very sensitive due to a super exponential dependency of the attenuation with
the wavelength in that range.
Using the correct EBL level to reconstruct the intrinsic spectrum,
the intrinsic spectrum is well described by a smooth function (Fig.~\ref{Fig:MidEnergyExampleOfMethod}, middle panel).
Instead, when using a "wrong" scaled EBL density characteristic deviations
(wiggles) from a smooth function are visible (Fig.~\ref{Fig:MidEnergyExampleOfMethod}, lower panel). 
The wiggles are quantified by a reduced $\chi^{2}$ value of 61.4/12, corresponding
to a fit probability of $1.3 \cdot 10^{-8}$.
The small fit probability for the assumed scaled EBL level
implies (under used assumptions) significant presence of the unphysical wiggles and,
therefore, the exclusion of that particular EBL realization.

\begin{figure}[tbp]
\centering
\includegraphics[width=0.48\textwidth]{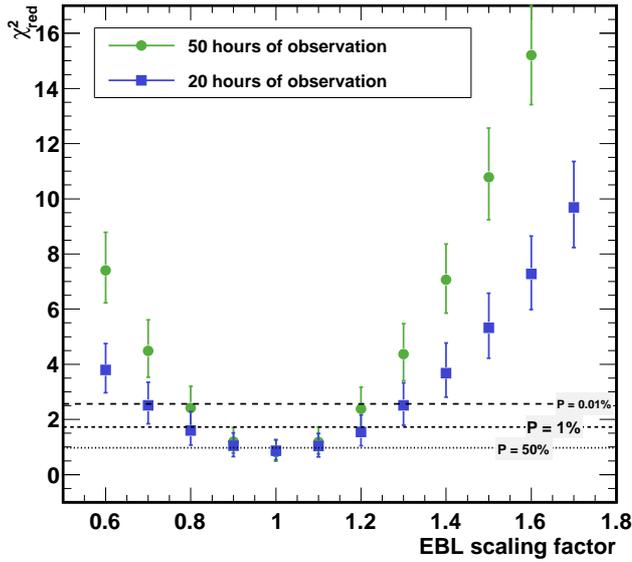}
\caption{
Quantitative results from the search for EBL signatures in the mid-VHE using
the energy spectrum of Mkn\,501. The results for the array layout $I$ are shown: 
the reduced $\chi^{2}$ values from
the fits to the reconstructed spectra of Mkn\,501 as a function of EBL scaling
factor. Blue filled squares and green filled circles show the expected result
for 20 and 50 hours of CTA observations, respectively. The black horizontal lines
correspond to fit probabilities as labeled.}
\label{Fig:MidEnergyStatistics}
\end{figure}

\paragraph{Results of the analysis}
This procedure is repeated 1000 times for every scaled EBL density in order
to achieve a solid statistical mean and a $1\,\sigma$ (68\%) coverage of the
reduced $\chi^{2}$ values. The results are shown in
Fig.~\ref{Fig:MidEnergyStatistics} for two different CTA exposures (20 and 50
hours) plotting the mean reduced $\chi^{2}$ values including 68\% error bars
versus the EBL scaling factor. As a result one can see a parabola-like curve
with the minimum at EBL scaling factor of  1.0, which is via construction the
correct EBL model.  The expected mean reduced $\chi^{2}$ values are shown by
the blue filled squares and the green filled circles for the exposures of
20\,hr and 50\,hr, respectively. The corresponding fit probabilities of P=50\%,
1\% and 0.01\% are shown by the dotted, short-dashed and long-dashed lines,
respectively.
An EBL scaling is considered to be excluded when the mean reduced $\chi^{2}$
value including its 68\% error exceeds the $\chi^{2}$ value for P=0.01\%.
For the case constructed here this means that in case of 50 hours observation,
the EBL scalings below 0.70 and above 1.30 are excluded. In the case of 20
hours observation, the EBL scalings factors above 1.40 and below 0.6 are excluded.

When comparing performance of different CTA array layouts for this particular study,
no significant differences between the results was found for the array layouts $B$, $I$, and $C$.
The reason for low discrepancies is that the study mainly relies on the energy resolution of the array
in the range 200 GeV up to few TeV and per construction this is the regime where all CTA arrays  
have similar performances.

%------------------
% FIGURE: Mid energies - statistics F11
 %------------------
%--------------------------------------------------------------------------------
% Conclusions, recommendations, & caveats
%--------------------------------------------------------------------------------
\section{Conclusions, recommendations, \& caveats}

CTA will be a superb machine for probing the EBL. Key features are the high
sensitivity, providing a large and rich sample of sources for EBL studies, and
the huge energy coverage, enabling to simultaneously measure the EBL
attenuation and the un-attenuated intrinsic spectrum. CTA will deliver definite
answers to some of the most pressing topics in this field, including a precise
measurement of the EBL density and it evolution up to redshifts of one and
beyond, providing new insights into structure formation and galaxy evolution.
In addition, it will open up the field of EBL studies towards more fundamental
physics, allowing to probe for effects of fundamental particles like hidden
photons or axions
(see  \cite{barrio:2011a}), and give constraints on cosmology.

The currently proposed layouts come in three different flavors: (1) 
emphasis on low energy sensitivity and lower threshold with worse high energy performance 
(array layout $B$), and (2) increased high energy performance with worse acceptance at low energies (array
layout $C$), and (3) overall
balanced sensitivity with emphasis on the core energy range between 100\,GeV and
10\,TeV (e.g. array layout $E$ or $I$). Each of the setups have their advantages for certain topics of the
EBL studies: arrays of type (1) are thought to be better suited for studies of UV radiation and early universe using 
GRBs and distant AGNs where the most observational emphasis is put on energies below 100 GeV; array layouts
of type (2) provide the opportunity to study in great detail the
mid and far IR with sources like
 M\,87 and 1ES\,0229+200 
 where VHE gamma ray emission extends
in energy up to 10s of TeV; and while array layouts of type (3) seem to deliver the best compromise
of the previous two types.

Several CTA array layouts have been investigated in detail for their performance for different EBL and cosmology physics cases. 
The results are:
\begin{itemize}
 \item For low energy studies of steady objects (i.e. implying long exposure), array of type $B$ is superior in its performance 
thanks to its large collection area and good background rejection at energies below 100 GeV. 
Due to their decreased sensitivity at lower energies, the capability of sampling the un-attenuated part of the spectrum with array layouts $I$ and $E$ is slightly worse (see Sec.\ref{Sec:DisentaglingIntrFromEBL}). Array layout $C$ shows significant disadvantages for these type of studies.
 \item For low energy studies of rapid transient phenomena like GRBs, no significant difference between the performance of arrays $B$ and $I$
could be found. Due to the short exposure, the background rejection does not
play an important role, equalizing the performance of the two types of arrays
(see Sec.~\ref{Sec:grbssection}). Instead, an array of type (3) (e.g. $C$) with a
significantly higher energy threshold does not allow for a detection of the
phenomena and is, therefore, unsuited for such studies.  
 \item For the precision measurements of the $z=0$ EBL
density as discussed in Sec.~\ref{Sec:PrecisionMeasurementOfZ0EBL}, all tested arrays show very similar performance. 
The EBL can be resolved down to 20-30\% thanks to the superb spectral resolution of the planned CTA. However, in order
to be sensitive to the differences between the state-of-the-art EBL models, an energy resolution of 5-10\% in the
energy range between 100 GeV and few TeV is required.
\end{itemize}

 \section*{Acknowledgement}
The authors thank the CTA consortium members for fruitful
discussions and the referees for insightful comments and
suggestions.

%--------------------------------------------------------------------------------
%References
%--------------------------------------------------------------------------------

%% References

%------------------------------------------------------------------------------------------------------------------------
% Bibliography
%------------------------------------------------------------------------------------------------------------------------

%\input{../../../../bibtex/journalNames.tex}
%\bibliographystyle{../../../aAndAStyles/a	a-package/bibtex/aa}

%\bibliographystyle{../../../styles/elsevier/elsarticle-num}
%\bibliography{../../../../bibtex/bibtex_db_2012}

\section*{References}

\bibliographystyle{elsarticle-num}
\bibliography{main}
%\bibliography{/Users/mazin/physics/bibtex/bibtex_db_2011}


%\bibliographystyle{elsarticle-num}





% \end{thebibliography}

\end{document}